\documentclass[fleqn,usenatbib]{mnras}

\usepackage[dvipdfmx]{graphicx}

\usepackage{amsmath}
\usepackage{color}

\usepackage[T1]{fontenc}
\usepackage{aecompl}
\usepackage{yfonts}

\begin{document}

\title[]{The relation between the mass-to-light ratio and the relaxation state of globular clusters}

\author[P. Bianchini et al.]{P. Bianchini$^{1,}$\thanks{E-mail:
bianchip@mcmaster.ca}\thanks{CITA National Fellow},
A. Sills$^{1}$,
G. van de Ven$^{2}$ \&
A. C. Sippel$^{2}$
\\
$^{1}$Department of Physics and Astronomy, McMaster University, Hamilton, Ontario, L8S 4M1, Canada\\
$^{2}$Max-Planck Institute for Astronomy, Koenigstuhl 17, 69117 Heidelberg, Germany\\
}

\date{}
\maketitle

\begin{abstract}
The internal dynamics of globular clusters (GCs) is strongly affected by two-body interactions that bring the systems to a state of partial energy equipartition. Using a set of Monte Carlo clusters simulations, we investigate the role of the onset of energy equipartition in shaping the mass-to-light ratio (M/L) in GCs. Our simulations show that the M/L profiles cannot be considered constant and their specific shape strongly depends on the dynamical age of the clusters. Dynamically younger clusters display a central peak up to M/L $\simeq25$ $M_\odot/L_\odot$ caused by the retention of dark remnants; this peak flattens out for dynamically older clusters. Moreover, we find that also the global values of M/L correlate with the dynamical state of a cluster quantified as either the number of relaxation times a system has experienced $n_{rel}$ or the equipartition parameter $m_{eq}$: clusters closer to full equipartition (higher $n_{rel}$ or lower $m_{eq}$) display a lower M/L. We show that the decrease of M/L is primarily driven by the dynamical ejection of dark remnants, rather than by the escape of low-mass stars. The predictions of our models are in good agreement with observations of GCs in the Milky Way and M31, indicating that differences in relaxation state alone can explain variations of M/L up to a factor of $\simeq3$. Our characterization of the M/L as a function of relaxation state is of primary relevance for the application and interpretation of dynamical models.
\end{abstract}


\section{Introduction}

Globular clusters (GCs) are old stellar systems that have been strongly shaped by the ongoing gravitational interactions between their stars. These gravitational encounters, together with the interactions with the host galaxy and stellar evolution processes, are responsible for GCs long term evolution and their current observable properties.

Amongst other properties, the current mass-to-light ratio (M/L) of GCs is of particular interest, since it links the stellar luminosity of a system to its total mass. The study of M/L and its evolution, not only provides direct information about the present stellar population of a GC and its present total gravitational potential, but can also be used as a tool for studies of the stellar initial mass function (IMF) (see e.g. \citealp{Zaritsky2014,Schaerer2011}). For this latter case, GCs are particularly useful since they can be approximated as single-age, single-metallicity and dark matter free systems, with the gravitational potential due to their stars only. 

The M/L in GCs depends on the initial mass function, metallicity, stellar evolution (therefore the age of the cluster) and the dynamical evolution of the cluster. Stellar evolution has the effect of increasing the M/L since the ratio between high-mass stars (characterized by low M/L) and low-mass stars (characterized by high M/L) decreases with time, while massive stars gradually evolve into stellar remnants. The dynamical evolution of the clusters is instead shaped by two-body relaxation that drives the energy exchange between stars, leading massive stars to lose kinetic energy and sink toward the centre, while less massive stars gain kinetic energy and move toward the outskirt. The system therefore becomes mass segregated and reaches a state of partial energy equipartition (see, e.g. the recent works of \citealp{TrentivanderMarel2013,Bianchini2016b,Spera2016}). Moreover, the onset of energy equipartition can lead to the preferential escape of low-mass stars (\citealp{Vesperini1997,BaumgardtMakino2003,Kruijssen2008}). Both these effects (mass segregation and preferential loss of low-mass stars) have direct implications on shaping the  M/L value of GCs. 

Using a set of $N$-body simulations, \citet{BaumgardtMakino2003} showed that the preferential depletion of low-mass stars from star clusters leads to a decrease of their M/L ratios except for a short period close to final dissolution. Similarly, using analytical models, \citet{Kruijssen2008} and \citet{KruijssenLamers2008} showed that dynamical evolution of clusters strongly affects their luminosity and M/L ratios and could explain the observed values of M/L and their increase as the mass of the cluster increases (see also \citealp{Goudfrooij2016}).

The evolution of the M/L for young star clusters has been studied by \citet{Boily2005} and \citet{Fleck2006}, showing how the application of standard dynamical models (e.g. virial estimates of the dynamical mass) can be strongly biased if the migration of massive stars toward the clusters centres is not taken into account. This work emphasized the importance of mass segregation in the application of dynamical modelling, that generally rely on simplifying assumptions, such as a constant and fixed M/L\footnote{For example, the Jeans model approach (e.g. \citealp{Watkins2015b}) or one-component distribution-function modelling (e.g. \citealp{MLvdM2005,Zocchi2012,Bianchini2013}).}. This assumption could limit the validity of the modelling techniques, resulting in the misinterpretation of the results. For this reason, recent works have focused on incorporating mass segregation as a physical ingredient in dynamical modelling techniques (\citealp{ShanahanGieles2015,GielesZocchi2015,Zocchi2016,deVita2016}).

Using a set of Monte Carlo cluster simulations, \citet{Bianchini2016b} showed that the onset of energy equipartition has a fundamental role in determining the internal kinematic properties of GCs, in particular in generating velocity dispersion that is mass-dependent, $\sigma(m)$, with $m$ the stellar mass. Despite the fact that only partial energy equipartition is reached (e.g., \citealp{TrentivanderMarel2013} and references therein), the study of this mass-dependent kinematics is now possible in observations thanks to state-of-the-art kinematic data (e.g. a combination of \textit{Hubble Space Telescope} proper motions, \citealp{Bellini2014}, and line-of-sight data, \citealp{Kamann2016}). These data will allow one to quantify the level of energy equipartition reached by a cluster using, for example, the parameter $m_{eq}$, introduced by \citet{Bianchini2016b}, uniquely describing the shape of the $\sigma(m)$ profile through the relation
\begin{equation}
 \sigma(m) = \sigma_0\,\exp\left(-\frac{1}{2}\frac{m}{m_\mathrm{eq}}\right).
\label{eq:1}
\end{equation}

This study showed that the level of energy equipartition in a GC strongly correlates with its dynamical state, quantified by the number of relaxation times experience by the cluster, $n_{rel}=T_{age}/T_{rc}$. This relation between the dynamical state and the degree of energy equipartition has proven fundamental to predict the mass dependence of the velocity dispersion in stellar systems for which the relaxation states were known (see \citealp{Baldwin2016, Bianchini2016a} for applications to studies related to blue stragglers stars and binary stars, respectively).
 
A first study on the connection between the effects of energy equipartition and mass segregation was put forward by \citet{WebbVesperini2016b} using a set of $N$-body simulations \citep{WebbVesperini2016a}, showing that, within the half-mass radius, the degree of energy equipartition correlates with the degree of mass segregation (quantified with the change of the slope of the stellar mass function with radius). We develop this further, exploring the connection between the level of energy equipartition reached by a cluster, its relaxation state and the corresponding M/L ratio, using the same set of simulations introduced in \citet{Bianchini2016b}. The goal of our work is to quantify the variation of the M/L ratios in GCs as a function of the relaxation states, provide an understanding of the expected M/L profiles in present day GCs and allow for a direct comparison to observations and for the application of dynamical modelling.

In Section \ref{sec:2} we summarize the set of simulations used in this work and the relevant physical quantities extracted for each snapshots. In Section \ref{sec:3} we analyze the shapes of the M/L radial profiles and the global values of M/L and connect them with the relaxation state of the clusters. The relation between relaxation state and evolution of M/L ratios extracted from our models, is directly compared to available data sets of GCs (Section \ref{sec:comparison}). Finally, in Section \ref{sec:discussion} and \ref{sec:conclusions} we discuss the implications of our results and report our conclusions.

\section{Simulations}
\label{sec:2}

\begin{table}
\begin{center}
\caption{\textbf{Initial conditions of our set of simulations.} The original name of the simulations from \citet{Downing2010} are given in parentheses. We report the initial binary fraction $f_\mathrm{binary}$, the initial ratio of the intrinsic 3-dimensional tidal to half-mass radius $r_t/r_m$, the initial number of particles $N$, the initial mass $M$ and the metallicity [Fe/H]. Simulations from \citet{Downing2010}, except Sim 7, 10low75-2M, from private communication of J. M. B. Downing.}
\tabcolsep=0.1cm

\begin{tabular}{lcccccc}
\hline\hline
&$f_\mathrm{binary}$&$r_t/r_m$&N&M [$M_\odot$]& [Fe/H]\\
\hline

Sim 1 (10low75)&10\%&75&$5\times10^5$&$3.62\times10^5$& $-$1.3\\
Sim 2 (50low75)&50\%&75&$5\times10^5$&$5.07\times10^5$& $-$1.3\\
Sim 3 (10low37)&10\%&37&$5\times10^5$&$3.62\times10^5$& $-$1.3\\
Sim 4 (50low37)&50\%&37&$5\times10^5$&$5.07\times10^5$& $-$1.3\\
Sim 5 (10low180)&10\%&180&$5\times10^5$&$3.63\times10^5$& $-$1.3\\
Sim 6 (50low180)&50\%&180&$5\times10^5$&$5.07\times10^5$& $-$1.3\\
Sim 7 (10low75-2M)&10\%&75&$20\times10^5$&$7.26\times10^5$& $-$1.3\\
Sim 1-sol (10sol75)&10\%&75&$5\times10^5$&$3.61\times10^5$& 0.0\\
Sim 2-sol (50sol75)&50\%&75&$5\times10^5$&$5.05\times10^5$& 0.0\\
Sim 3-sol (10sol37)&10\%&37&$5\times10^5$&$3.63\times10^5$& 0.0\\
Sim 4-sol (50sol37)&50\%&37&$5\times10^5$&$5.08\times10^5$& 0.0\\
Sim 5-sol (10sol180)&10\%&180&$5\times10^5$&$3.61\times10^5$& 0.0\\
Sim 6-sol (50sol180)&50\%&180&$5\times10^5$&$5.07\times10^5$& 0.0\\
\hline

\end{tabular}

\label{tab:initial}
\end{center}
\end{table}

\begin{table*}
\tabcolsep=0.15cm
\begin{center}
\caption{\textbf{Projected properties of the set of simulations for the 4, 7, 11 Gyr snapshots.} The initial conditions of the simulations were reported in \citet{Bianchini2016b}. Here we report the concentration $c=\log(R_t/R_c)$, with $R_t$ and $R_c$ as projected tidal radius and projected core radius respectively, the half light radius $R_h$ in parsec, core radius $R_c$ in parsec, the logarithm of the half-light relaxation time  $T_\mathrm{rh}$ in yr, and the logarithm of the core relaxation time $T_\mathrm{rc}$ in yr. All simulations have an initial number of particles of N=500\,000, except for simulation 7 with N=2\,000\,000.}
\begin{tabular}{lrclrclrclrclrcl}
\hline\hline
 & \multicolumn{3}{c}{c} & \multicolumn{3}{c}{R$_h$} & \multicolumn{3}{c}{R$_c$} & \multicolumn{3}{c}{$\log T_\mathrm{rh}$} & \multicolumn{3}{c}{$\log T_\mathrm{rc}$}\\
&4 Gyr &7 Gyr &11 Gyr &4 Gyr &7 Gyr &11 Gyr &4 Gyr &7 Gyr &11 Gyr &4 Gyr &7 Gyr &11 Gyr &4 Gyr &7 Gyr &11 Gyr\\
\hline
Sim 1&1.52& 1.46& 1.45& 4.01&4.23&4.92& 2.74&3.12&3.15&9.382&9.487& 9.543&9.151&9.172&9.123\\
Sim  2&1.42&1.38 & 1.34& 4.89& 5.92&6.06& 3.42&3.62&3.89&9.474&9.579&9.655&9.345&9.287&9.286\\

Sim  3&1.26&1.21 & 1.16& 7.04& 8.16&9.05&4.92 & 5.52&6.07&9.658&9.755&9.820&9.647&9.656&9.645\\
Sim  4&1.21&1.16 & 1.12& 8.84&8.96&10.92&5.54 & 6.11&6.47&9.705&9.803&9.877&9.757&9.776&9.744\\

Sim  5&1.81& 1.95& 2.06& 1.53&1.90&2.69& 1.33&0.96&0.75&9.171&9.263&9.349&8.437&8.033&7.740\\
Sim  6&1.73&1.74 & 1.79& 2.96&3.10&3.05& 1.64&1.56&1.34&9.249&9.347&9.417&8.598&8.472&8.262\\

Sim 7 &1.52&1.52&1.51&2.57&2.62&2.90&1.73&1.87&1.85&9.415&9.498&9.565&9.040&8.965&8.991\\

Sim 1-sol & $-$ & $-$& 1.75&$-$ &$-$ &3.02 & $-$&$-$ & 1.59&$-$ &$-$ &9.505 & $-$&$-$ &8.445 \\
Sim 2-sol & $-$ & $-$& 1.55 &$-$ &$-$ &4.65 & $-$&$-$ & 2.48 &$-$ &$-$ &9.576 & $-$&$-$ & 8.866\\
Sim 3-sol & $-$ & $-$& 1.31 &$-$ &$-$ &5.88 & $-$&$-$ & 4.26&$-$ &$-$ &9.729 & $-$&$-$ &9.324 \\
Sim 4-sol & $-$ & $-$& 1.23&$-$ &$-$ &7.47 & $-$&$-$ & 5.20&$-$ &$-$ &9.781 & $-$&$-$ &9.510 \\
Sim 5-sol  & $-$ & $-$& 2.22&$-$ &$-$ & 1.72& $-$&$-$ & 0.51 &$-$ &$-$ &9.364 & $-$&$-$ & 7.265\\
Sim 6-sol  & $-$ & $-$& 1.87&$-$ &$-$ & 2.02& $-$&$-$ & 1.13 &$-$ &$-$ &9.407 & $-$&$-$ &8.127 \\

\hline
\end{tabular}
\label{tab:1}
\end{center}
\end{table*}
\begin{table*}
\tabcolsep=0.15cm
\begin{center}
\caption{\textbf{Relaxation states and mass-to-light ratios for the 4, 7, 11 Gyr snapshots of our simulations.} Number of core relaxation times a cluster has experienced $n_{rel}=T_\mathrm{age}/T_\mathrm{rc}$, equipartition parameter $m_\mathrm{eq}$ in $M_\odot$ indicating the state of partial energy equipartition reached as in \citet{Bianchini2016b}, global $V$-band mass-to-light ratio M/L, within the half-light radius $R_h$ and within $0.1R_h$, in solar units.}
\begin{tabular}{lrclrclrclrclrcl}
\hline\hline
 & \multicolumn{3}{c}{$n_{rel}=T_\mathrm{age}/T_\mathrm{rc}$}  & \multicolumn{3}{c}{$m_{eq}$} &\multicolumn{3}{c}{M/L global} & \multicolumn{3}{c}{M/L$(<R_h)$}  & \multicolumn{3}{c}{M/L$(<0.1R_h)$}\\
&4 Gyr &7 Gyr &11 Gyr &4 Gyr &7 Gyr &11 Gyr &4 Gyr &7 Gyr &11 Gyr &4 Gyr &7 Gyr &11 Gyr &4 Gyr &7 Gyr &11 Gyr\\
\hline
Sim 1	&2.8& 4.4& 8.3			 & 3.24&2.37&	2.00		& 1.04& 1.61&2.22		&0.85& 1.23& 1.84		&0.92&2.09&2.17\\
Sim  2	&1.8& 3.6& 5.7			& 4.18& 3.36&	2.73		&1.15 &1.95&2.95		&0.98& 1.72& 2.51 		&1.67&2.48&3.81\\

Sim  3	&0.9& 1.5& 2.5			 & 5.03& 3.84&	2.92		&1.19 & 1.66 & 2.59		&1.08& 1.52&2.35 		&1.69&2.59&2.82\\
Sim  4	&0.7& 1.2& 2.0			& 7.63& 5.58&	4.46		&1.38 & 2.05 &	3.26	 	&1.34& 1.88&3.12 		&2.58&4.80&5.60\\

Sim  5	&14.6& 64.9& 200.2		& 1.71&1.50&1.66		& 0.85& 1.22&1.78		&0.44& 0.69& 1.29 		&0.43&0.30&1.44\\
Sim  6	&10.1& 23.6& 60.1		& 2.48&2.04&1.65		&1.05 &1.55&2.20		&0.80& 1.15& 1.61 		&0.84&1.09&0.81\\

Sim 7 	&3.6& 7.6& 11.2		&3.07&2.19&1.90		&0.73& 1.00&1.40		&0.55& 0.70& 1.02	 	&0.54&0.59&0.99\\

Sim 1-sol  & $-$ & $-$& 39.5&$-$ &$-$ & 1.57 & $-$&$-$ &3.53 &$-$ &$-$ & 2.17& $-$&$-$ & 1.61\\
Sim 2-sol  & $-$ & $-$&14.9 &$-$ &$-$ & 2.26 & $-$&$-$ & 3.73&$-$ &$-$ & 3.02& $-$&$-$ & 2.73 \\
Sim 3-sol  & $-$ & $-$&5.2&$-$ &$-$ & 2.12  & $-$&$-$ & 3.62&$-$ &$-$ &2.69 & $-$&$-$ & 4.48\\
Sim 4-sol  & $-$ & $-$&3.4 &$-$ &$-$ & 3.11 & $-$&$-$ & 4.35&$-$ &$-$ &3.51 & $-$&$-$ & 3.84 \\
Sim 5-sol  & $-$ & $-$&597.9 &$-$ &$-$ & 1.50  & $-$&$-$ &3.47 &$-$ &$-$ & 1.82& $-$&$-$ &1.50 \\
Sim 6-sol  & $-$ & $-$& 82.1&$-$ &$-$ & 1.49 & $-$&$-$ &3.28 &$-$ &$-$ & 1.79& $-$&$-$ & 1.35 \\
\hline
\end{tabular}
\label{tab:2}
\end{center}
\end{table*}

We consider a set of Monte Carlo cluster simulations developed by \citet{Downing2010} with the Monte Carlo code of \citet{Giersz1998} (see also \citealp{Hypki2013}). This set of models was used in \citet{Bianchini2016b,Bianchini2016a}. In addition to these simulations, we consider a set of models characterized by the same initial conditions but differing only in metallicity, as described below.
The simulations all include an initial mass function, stellar evolution, primordial binaries, and a high number of particles, providing a realistic description of the long-term evolution of non-rotating GCs with a single stellar population.

\subsection{Initial conditions and observational properties}
The initial conditions are drawn from a \citet{Plummer1911} model, a \citet{Kroupa2001} initial mass function (with stellar masses between 0.1-150 $M_\odot$), a metallicity of either [Fe/H]=$-$1.3 (indicative of typical metal poor GCs) or  [Fe/H]=0 (indicative of typical metal rich GCs), and an initial tidal cut-off at 150 pc (making the simulations relatively isolated, comparable to halo clusters at $9-10$ kpc from the centre of the Milky Way). We consider a total of 12 simulations with 500\,000 initial particles, characterized by 3 values of initial concentrations, 2 values for the initial binary fraction (either 10\% or 50\%) and 2 values of metallicity. Additionally, we include a simulation with 2\,000\,000 particles and 10\% initial binary fraction. Table \ref{tab:initial} summarizes the initial conditions of the set of simulations.
All simulations assign a natal velocity kick to black holes and neutron stars, due to asymmetric supernova explosions, drawn from a Maxwellian velocity distribution with a peak at 190~km~s$^{-1}$ \citep{Downing2011}.\footnote{We note that the peak of the natal velocity kick distribution is well above the escape velocity of any of our simulations ($<50$ km s$^{-1}$ at the half-mass radius, \citealp{Downing2011}).} For each simulation we analyze the snapshots at 4, 7 and 11 Gyr, to assess the time evolution, with the exception of the metal rich simulations for which we investigate only the snapshots at 11 Gyr.

Table \ref{tab:1} summarizes the observational properties of the simulations (for the time-snapshots at 4, 7, 11 Gyr) as calculated in \citet{Bianchini2016b}. We report the concentration $c$ defined as $c=\log (R_t/R_c)$, with $R_t$ the projected tidal radius and $R_c$ the projected core radius; the projected core radius $R_c$, defined as the radius where the surface density is half of the central surface density\footnote{Calculated from number-count surface density profiles.}; the projected half-light radius $R_h$, containing half of the luminosity of the cluster; the logarithm of the half-mass relaxation time $T_\mathrm{rh}$ and the logarithm of the core relaxation time $T_\mathrm{rc}$ (see eq. 1 and 2 in \citealp{Bianchini2016b}). 

\subsection{Relaxation state}
\label{sec:ML}
In order to characterize the relaxation state of our set of simulations we compute for every snapshots the number of relaxation times a cluster has experienced $n_{rel}=T_{age}/T_{rc}$, defined as the ratio between the age of the cluster and its current core relaxation time. Our simulations span a large range in relaxation states, with the least relaxed simulation (dynamically young) characterized by $n_{rel}=0.7$ and the most relaxed simulation (dynamically old) characterized by $n_{rel}\approx600$ (see Table \ref{tab:2}).

The relaxation state of the cluster can also be characterized by the equipartition parameter $m_{eq}$ introduced by \citet{Bianchini2016b}, measuring the state of partial equipartition reached by a cluster. This parameter is derived solely from the internal kinematics, in particular from a fit of equation (\ref{eq:1}) to the relation between velocity dispersion and stellar mass (see also Section 3 of \citealp{Bianchini2016b}) and strongly correlates with the relaxation state  parameter $n_{rel}$. Dynamically old clusters show a stronger mass-dependence of the velocity dispersion and therefore are closer to full energy equipartition (characterized by a smaller $m_{eq}$); dynamically young clusters show only a weak mass-dependence of the velocity dispersion and therefore are far from a state of full energy equipartition (characterized by a larger $m_{eq}$). All our simulations are in a state of partial energy equipartition and we report in Table~\ref{tab:2} the corresponding values of $m_{eq}$.

Additionally, given our goal to link the relaxation state of a cluster to its mass-to-light ratio, we report in Table \ref{tab:2} the values of the mass-to-light ratios M/L of the different snapshots calculated summing the mass of the particles in a simulation (both stars and stellar remnants) divided by the corresponding $V$-band luminosity, from the output of the simulations. In particular we compute a global value of M/L using all particles in a simulation, a M/L$(<R_h)$ within the projected half-light radius and a central M/L$(<0.1R_h)$ within 0.1 half-light radius.
Note that these M/L ratios would correspond to dynamical mass-to-light ratios tracing the entire mass present in the cluster, not only the one of the luminous component\footnote{For example corresponding to mass-to-light ratios from virial estimates.}.

\begin{figure*}
\centering
\includegraphics[width=1\textwidth]{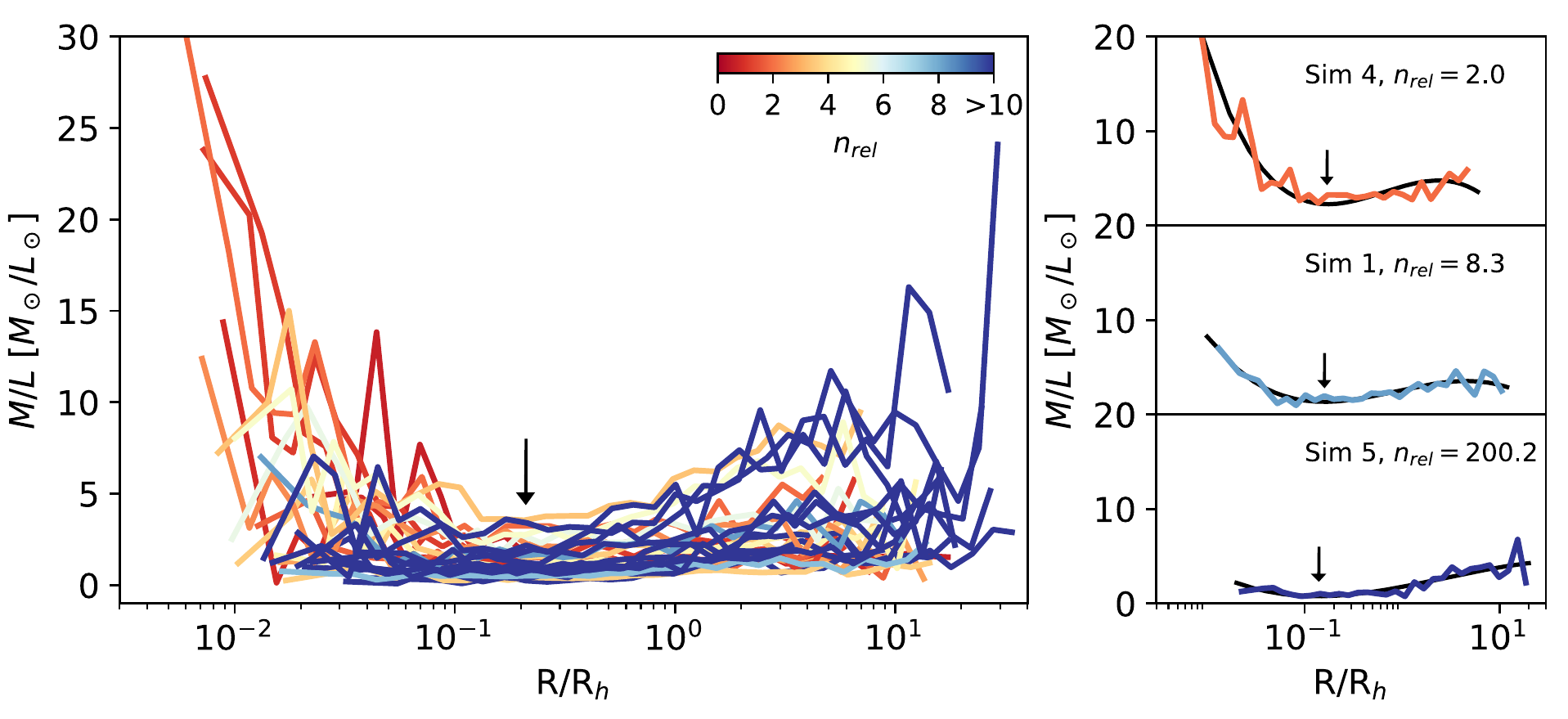}
\caption{\textbf{Left panel:} Mass-to-light ratio profiles (in the $V$-band) as a function of radius (in unit of the half-light radius, $R_h$) for all simulations. The profiles are colour-coded proportionally to their relaxation states indicated by $n_{rel}=T_{age}/T_{rc}$, see Table \ref{tab:2}. Redder profiles correspond to dynamically younger clusters while bluer profile dynamically older ones. \textbf{Right panels:} snapshots at 11 Gyr of three clusters in  different relaxation states ($n_{rel}=$2.0, 8.3, 200.2; see Appendix \ref{appA} for all the other profiles). Less relaxed clusters display a central peak of the mass-to-light ratio due to the presence of dark remnants not yet dynamically ejected from the clusters (see Sect. \ref{sec:darkremn} and Fig. \ref{fig:remn_vs_MS}). A polynomial fit is provided for each profile (see Appendix \ref{appA}) and is used to determine the minimum of the mass-to-light profiles, here indicated with a black arrow. A common minimum at $\simeq0.2-0.3 R_h$ is observed.}
\label{fig:ML_prof}
\end{figure*}

\section{Dependence of M/L on the relaxation states}
\label{sec:3}

In this section we investigate the connection between the relaxation state of our cluster simulations and their mass-to-light ratio. First we characterize in detail the radial variation of the M/L profile and its dependence on the dynamical state of a cluster, then we focus on global properties that can be directly compared to observations (e.g., global M/L or M/L within the half-light radius).

\subsection{M/L profiles}
\label{sec:darkremn}

We construct the mass-to-light ratio profiles for all the snapshots, using a radial binning with logarithm spacing. The mass-to-light ratio is computed considering all the objects (stars and stellar remnants) in each spatial bin, and considering the $V$-band luminosity. The profiles are shown in Fig. \ref{fig:ML_prof} as a function of radius, in units of the corresponding half-light radius, and are colour-coded according to their relaxation state $n_{rel}=T_{age}/T_{rc}$, reported in Table \ref{tab:2}.

From Fig. \ref{fig:ML_prof} we observe a clear dependence of the shape of the mass-to-light profile of a cluster on its relaxation state. Less relaxed clusters (lower $n_{rel}$, red colours) display a central peak (reaching values of M/L up to $\simeq25$~$M_\odot/L_\odot$, within $0.1R_h$) that progressively flattens for more relaxed snapshots (higher $n_{rel}$, blue colours). For the most relaxed clusters a slight increase in M/L is observable in the outer part (see right panels of Fig. \ref{fig:ML_prof}). As expected for mass-segregated systems, we note that the profiles cannot be considered constant with radius, not even in the cases for dynamically young snapshots. In fact, systems with low $n_{rel}$ (with ages of the same order of their relaxation time) are the ones that exhibit the strongest radial variation, namely a central peak.

The shape of the profiles is fully consistent with the average shape of M/L profiles reported by \citet{Baumgardt2017} from $N$-body simulations applied to Milky Way GCs (their figure 3). In Appendix \ref{appA} we provide polynomial fits to the M/L profiles and use these to determine the corresponding minima.
As noted by \citet{Baumgardt2017}, the presence of a minimum is due to the gradual mass segregation of stars with low M/L (giant stars and high-mass main sequence stars) towards the cluster centre, while the increase in the outer parts is due to the presence of low-mass main sequence stars. Interestingly, the profiles show a common minimum around $\simeq0.2-0.3R_h$, as indicated by a black arrow in Fig. \ref{fig:ML_prof} (see also Appendix~\ref{appA}).

\begin{figure*}
\centering
\includegraphics[width=1\textwidth]{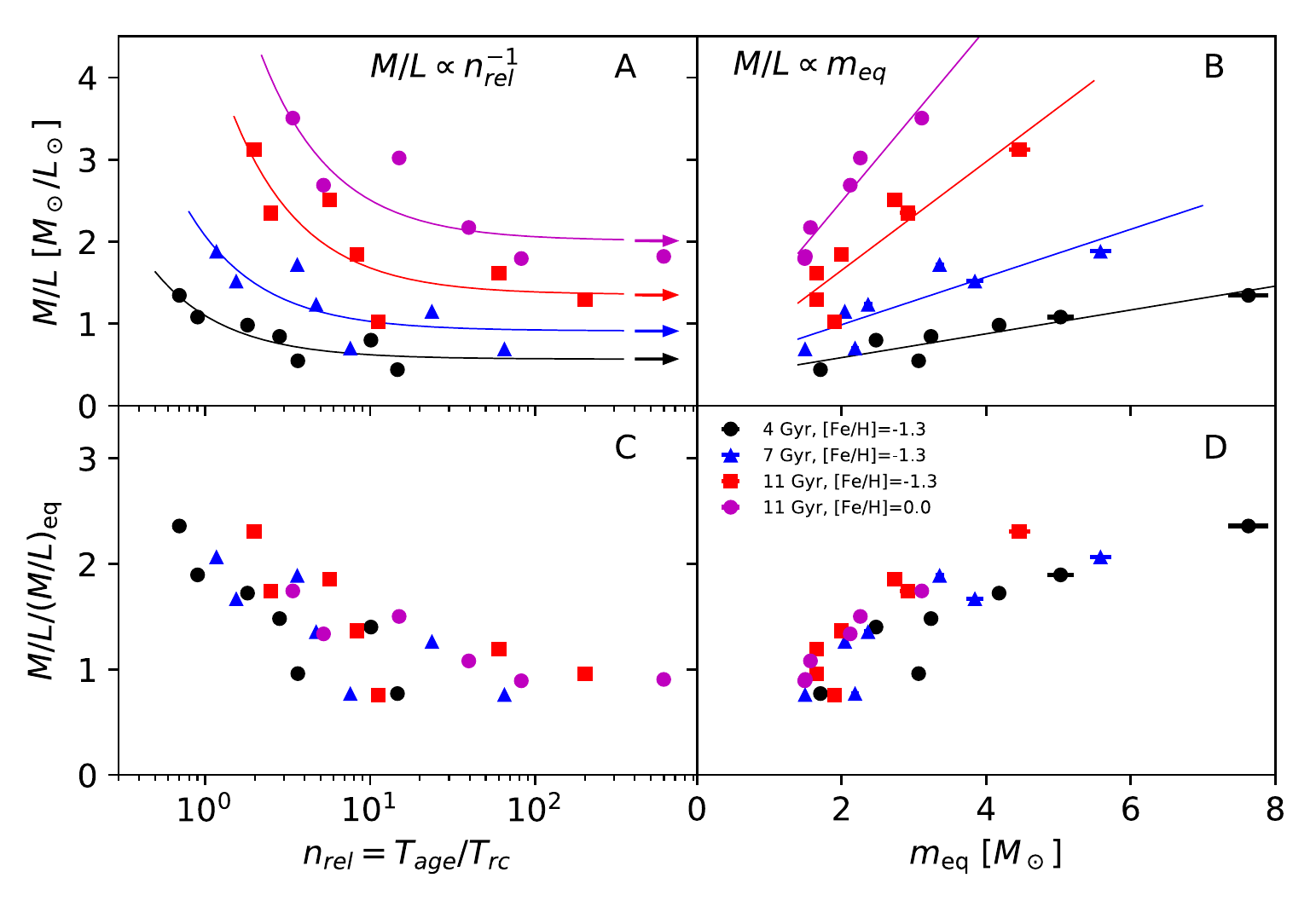}
\caption{\textbf{Top panels:} Mass-to-light ratio ($V$-band) within the half-light radius as a function of the relaxation state of a cluster $n_{rel}=T_{age}/T_{rc}$ (panel A) and its state of partial energy equipartition $m_{eq}$ (panel B). Metal poor models at 4, 7 and 11 Gyr are indicated with black circles, blue triangles and red squares, respectively, and metal rich models at 11 Gyr with magenta circles. More relaxed clusters (characterized by a higher $n_{rel}$, and, correspondingly, lower $m_{eq}$) have lower mass-to-light ratios. The arrows indicate the values of M/L reached by the clusters closest to full energy equipartition. A clear age dependence is observable, in agreement with stellar evolution (older stellar systems are depleted of brighter main sequence stars), as well as a metallicity dependence (metal rich clusters have lower luminosity due to higher stellar opacity). The solid lines are the best fits to each age-sequence. \textbf{Bottom panels:} Normalized mass-to-light ratio within the half-light radius as a function of the relaxation state of a cluster and its state of partial energy equipartition. Given an age and metallicity, the M/L are normalized with respect to the value that a cluster would reach at a very dynamically advanced state, $n_{rel}>>1$ (closer to full energy equipartition, i.e. the asymptotic limit of the fitted relation $M/L=a\, n_{rel}^{-1}+b$, see equation 3), indicated by arrows in panel A. This normalization allows to cancel out the effects connected to stellar evolution and metallicity and confirms that the correlation between relaxation state and M/L is due to dynamical effects alone.}
\label{fig:MLh_nrelc}
\end{figure*}

\begin{figure*}
\centering
\includegraphics[width=0.95\textwidth]{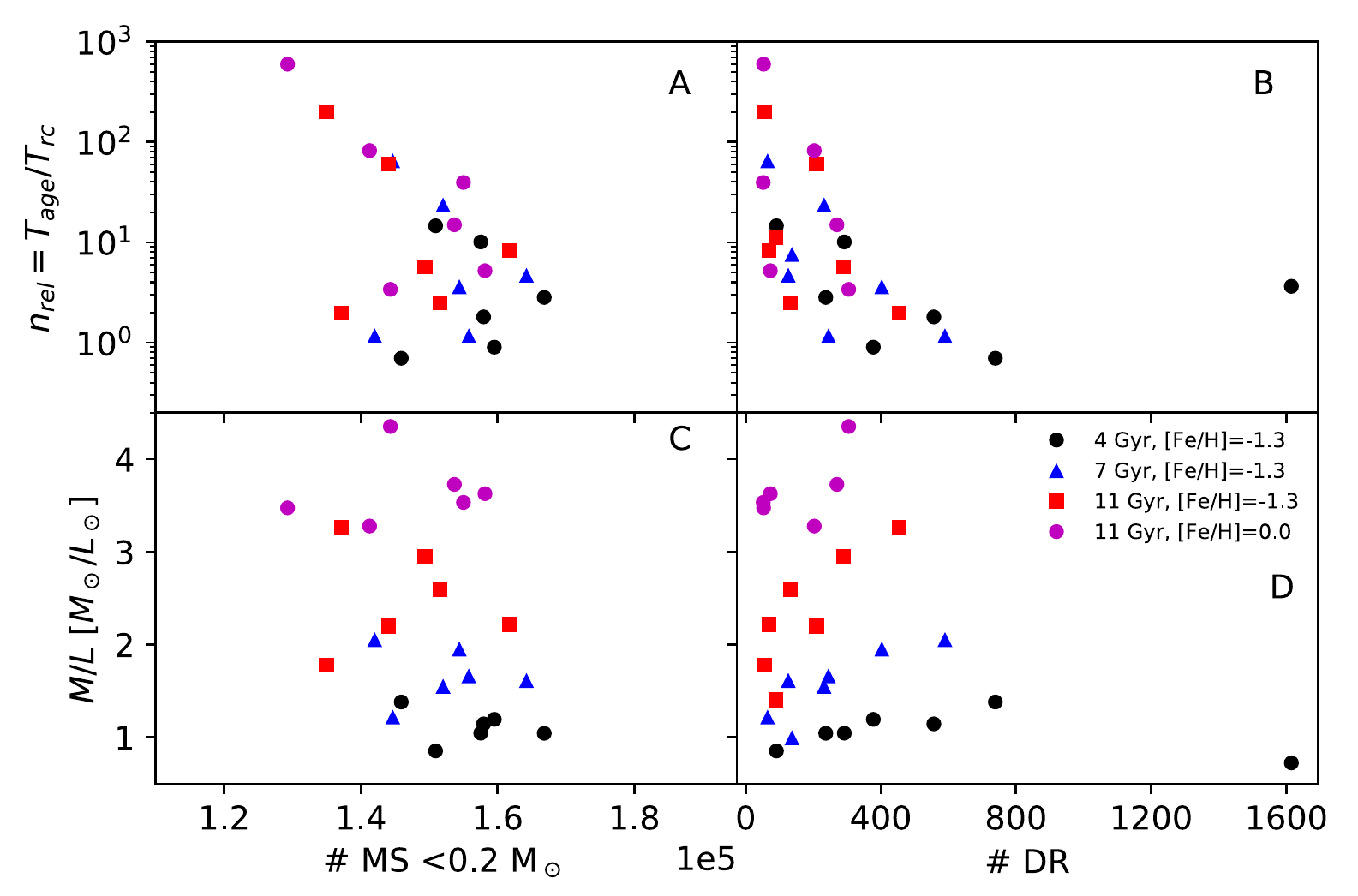}
\caption{\textbf{Top panels:} dependence of the total number of low-mass main sequence stars (MS, $m<0.2$ $M_\odot$; panel A) and dark remnants (DR, neutron stars and stellar mass black holes; panel B) on the relaxation state of a cluster $n_{rel}=T_{age}/T_{rc}$ (\textit{left panel}). Metal poor models at 4, 7 and 11 Gyr are indicated with black circles, blue triangles and red squares, respectively, and metal rich models at 11 Gyr with magenta circles. More relaxed clusters show a lower number of dark remnants because of ejections driven by gravitational encounters. No clear trend is observed for low-mass main sequence stars.
\textbf{Bottom panels:} relation between the total number of low-mass main sequence stars (panel C) and dark remnants (panel D) and the global mass-to-light ratio in the $V$-band. The strong correlation between the number of dark remnants and M/L, indicates that the dynamical evolution of  M/L is primarily driven by the ejection of dark remnants, rather than the escape of low-mass stars.}
\label{fig:remn_vs_MS}
\end{figure*}

To explain the central peak in the M/L profiles (within $0.1R_h$) and its relation with the relaxation states of the clusters, we need to investigate the combined role of dark remnants (high-M/L neutron stars and stellar black holes) and bright stars (low-M/L red giant stars).
Although both red giants and stellar remnants segregate toward the centre of a cluster, they are responsible for opposite variations of the M/L: dark remnants enhance the M/L, while bright red giants lower it. We compute the fraction of dark remnants to red giants stars within the central $0.1R_h$ and its variation with respect to the relaxation states of the clusters, $n_{rel}$. Less relaxed clusters have a higher fraction of dark remnants than more relaxed clusters in their core. This is due to the fact that, while a cluster reaches a more advanced state of dynamical evolution, the number of stellar encounters that the dark remnants have experienced is higher, and therefore they have experienced a higher number of dynamical ejections. \citet{Hurley2016} showed that the number of dark remnants (in particular black holes) remaining in a cluster depends primary on the density of the cluster, that is directly related to the frequency of gravitational encounters that the dark remnants experience \citep{Hills1992}. At a fixed age, lower density clusters (i.e, clusters with longer relaxation times) will retain a higher number of dark remnants for longer times \citep{Morscher2013}. This is also discussed in Section \ref{sec:globalML} (see panel B of Fig. \ref{fig:remn_vs_MS}).

Moreover, given their higher masses, dark remnants sink faster toward the centre and can quickly form a central component and enhance the escape rate of bright stars sinking in the centre (see e.g., \citealp{Trenti2010, Luetzgendorf2013}). This can further deplete the centre of the system of low-M/L stars.  

We can therefore explain the central peak observed in dynamically young clusters (see Fig. \ref{fig:ML_prof}) as due to a higher retention of dark remnants and fewer centrally segregated giant stars. This peak is progressively flattened for dynamically older clusters, due to the ejection of dark remnants. Note that at the later age of 11 Gyr, both dynamically young and dynamically old clusters are present ($n_{rel}$ ranges from $\simeq2$ to $\simeq600$, see Table \ref{tab:2}) and therefore both M/L profiles with and without central peak are expected. This has direct implications for the application of dynamical modelling when assuming a constant M/L. Knowing the relaxation state of a cluster, can help to establish what shape to expect for the M/L profile, and therefore helping to lower the degeneracies (such as mass-anisotropy degeneracy) in the models and disentangle the dynamical effects of stellar dark remnants from those of other dark components (e.g. intermediate-mass black holes or dark matter halos). This is already within our observational capabilities (see e.g. \citealp{Baldwin2016}), since the relaxation state of a cluster can either be estimated using the relaxation time already available for MW GCs and some extragalactic clusters (see Section \ref{sec:comparison}), or independently, using kinematic observations of mass dependent velocity dispersion, available from \textit{HST} proper motions \citep{Bellini2014} and line-of-sight data \citep{Kamann2016}.

\subsection{Global relations}
\label{sec:globalML}

With the goal of allowing for a direct comparison with observations, in this section we study the relation between the global values of mass-to-light ratios and the dynamical state of GCs. As described in Section \ref{sec:ML} and reported in Table \ref{tab:2}, we calculate the global values of M/L and the one within the half-light radius. In real observations, these quantities are generally the output of the application of dynamical modelling, and are usually referred to as dynamical mass-to-light ratios.

In Fig. \ref{fig:MLh_nrelc} we show the relation between the M/L within the half-light radius\footnote{Similar relations hold also for the global value of M/L and M/L within $0.1R_h$.}, the relaxation state $n_{rel}$ and the state of partial energy equipartition $m_{eq}$ reached by a cluster. We remind the reader that the state of partial energy equipartition of a cluster quantified by the parameter $m_{eq}$ tightly correlates with its relaxation state 
\begin{equation}
\label{eq:nrel_meq}
m_{eq}=1.55+4.10\,n_{rel}^{-\alpha},
\end{equation}
with $\alpha=0.85\pm0.12$, as described in \citet{Bianchini2016b} (see their Section 5). The equipartition parameter $m_{eq}$, solely calculated from the internal kinematics, can therefore be used to describe the dynamical age of a cluster.

Fig. \ref{fig:MLh_nrelc} shows that more relaxed clusters are characterized by a lower mass-to-light ratio. Given an age, the variation of M/L between dynamically younger and dynamically older clusters can be up to a factor of $\simeq3$. As an example, in the case of the metal poor models at 11 Gyr, the least relaxed cluster ($n_{rel}=2.0$; $m_{eq}=4.46$ $M_{\odot}$) has a M/L$(<R_h)=3.12$ $M_{\odot}/L_{\odot}$ and the most relaxed ($n_{rel}=200.2$; $m_{eq}=1.60$ $M_{\odot}$) has a M/L$(<R_h)=1.29$~$M_{\odot}/L_{\odot}$.

The relations between the M/L and the relaxation state ($n_{rel}$ and $m_{eq}$) presented in Fig. \ref{fig:MLh_nrelc} are tight. The M/L$-n_{rel}$ relations (panel A) are well fitted by a power law
\begin{equation}
\label{eq:nrel}
M/L=a\, n_{rel}^{-1}+b,
\end{equation}
with $a=0.53$, 1.16, 3.25, 4.97 and $b=0.57$, 0.91, 1.35, 2.01 for 4, 7, 11~Gyr metal poor and 11~Gyr metal rich clusters, respectively. While the M/L$-m_{eq}$ relations (panel~B) are well fitted by a linear relation
\begin{equation}
\label{eq:meq}
M/L=a\, m_{eq}+b,
\end{equation}
with  $a=0.15$, 0.29, 0.66, 1.05 and $b=0.30$, 0.41, 0.33, 0.39 for 4, 7, 11~Gyr  metal poor and 11~Gyr and metal rich clusters, respectively. Equations \ref{eq:nrel} and \ref{eq:meq} together give $m_{eq}\propto n_{rel}^{-1}$, consistent with what found in \citet{Bianchini2016b}; see equation \ref{eq:nrel_meq}.

The age dependence observed in panel A and B of Fig.~\ref{fig:MLh_nrelc} is due to stellar evolution effects (e.g., \citealp{Sippel2012, Zonoozi2016}): in older clusters the brighter and more massive stars gradually evolve into stellar remnants (white dwarfs, neutron stars and black holes) and therefore the mass-to-light ratio increases. Moreover, at a fixed age of 11 Gyr, metal rich clusters have on average higher M/L, because of a decrease in luminosity due to higher stellar opacity (e.g. \citealp{Sippel2012}), as expected from stellar population models.
In panel C and D of Fig. \ref{fig:MLh_nrelc}, we show the same relations after normalizing the M/L with respect to the M/L reached by a cluster fully dynamically relaxed, $n_{rel}>>1$, corresponding to a cluster at its closest state of full energy equipartition, $M/L_{eq}$. Using equations (3) and (2), $M/L_{eq}=0.57$, 0.91, 1.35, 2.01 $M_\odot/L_\odot$ and $m_{eq}=1.55$ $M_\odot$. This normalization allows to cancel out any age and metallicity effects, as seen from the smooth relations obtained, leaving only the effects of dynamical evolution.

Fig. \ref{fig:MLh_nrelc} and the resulting relations given by equations \ref{eq:nrel} and \ref{eq:meq}, show that the relaxation state of the cluster (and therefore its relaxation time) is responsible for shaping the mass-to-light ratios. This suggests that, on one side, the onset of energy equipartition and mass segregation (tightly related to the relaxation state of a cluster) could trigger the preferential escape of low-mass stars and on the other side, the relaxation state (that can be seen as a measure of how many gravitational encounters an object has experienced) regulates the retention/ejection of dark remnants.

In Fig. \ref{fig:remn_vs_MS}, we study which effect is stronger in lowering the mass-to-light ratios while the cluster dynamically evolves: the ejection of dark remnants or the escape of low-mass stars. Panels A and B of Fig. \ref{fig:remn_vs_MS} show the relaxation state $n_{rel}$ of the cluster as a function of the total number of main sequence stars with mass lower than 0.2 $M_\odot$ and the total number of dark remnants (neutron stars and stellar black holes). While low-mass main sequence stars do not show any clear trend with relaxation state\footnote{This is valid also for main sequence stars with masses $0.2<m<0.3$ $M_\odot$ and $0.3<m<0.4$ $M_\odot$.}, a strong anti-correlation is observable for the dark remnants (as already noted in Sect. \ref{sec:darkremn}).

Panels C and D of Fig. \ref{fig:remn_vs_MS} further show the global M/L ratios as a function of the total number of low-mass main sequence stars and the total number of dark remnants. The number of low-mass stars does not affect the M/L, while a strong correlation between M/L and number of dark remnants is observable.
We conclude that the dynamical lowering of M/L in our simulations is primarily driven by the dynamical ejection of stellar remnants, rather than by the preferential escape of low-mass stars. This result supports the finding by \citet{Sippel2016} that showed that the change of integrated colour of GCs is not driven by the removal of low-mass stars, but rather by the presence of the brightest stars. 
The dependence of the M/L on the number of dark remnants retained, opens up the possibility of constraining the number of dark remnants present in a clusters based on accurate measurements of the dynamical mass-to-light ratio (see also \citealp{Peuten2016}, for a kinematic approach to constrain the number of stellar-mass black holes).

Finally we note that the values of M/L are not directly effected by the primordial binary fractions of our simulations.

\begin{figure*}
\centering
\includegraphics[width=1\textwidth]{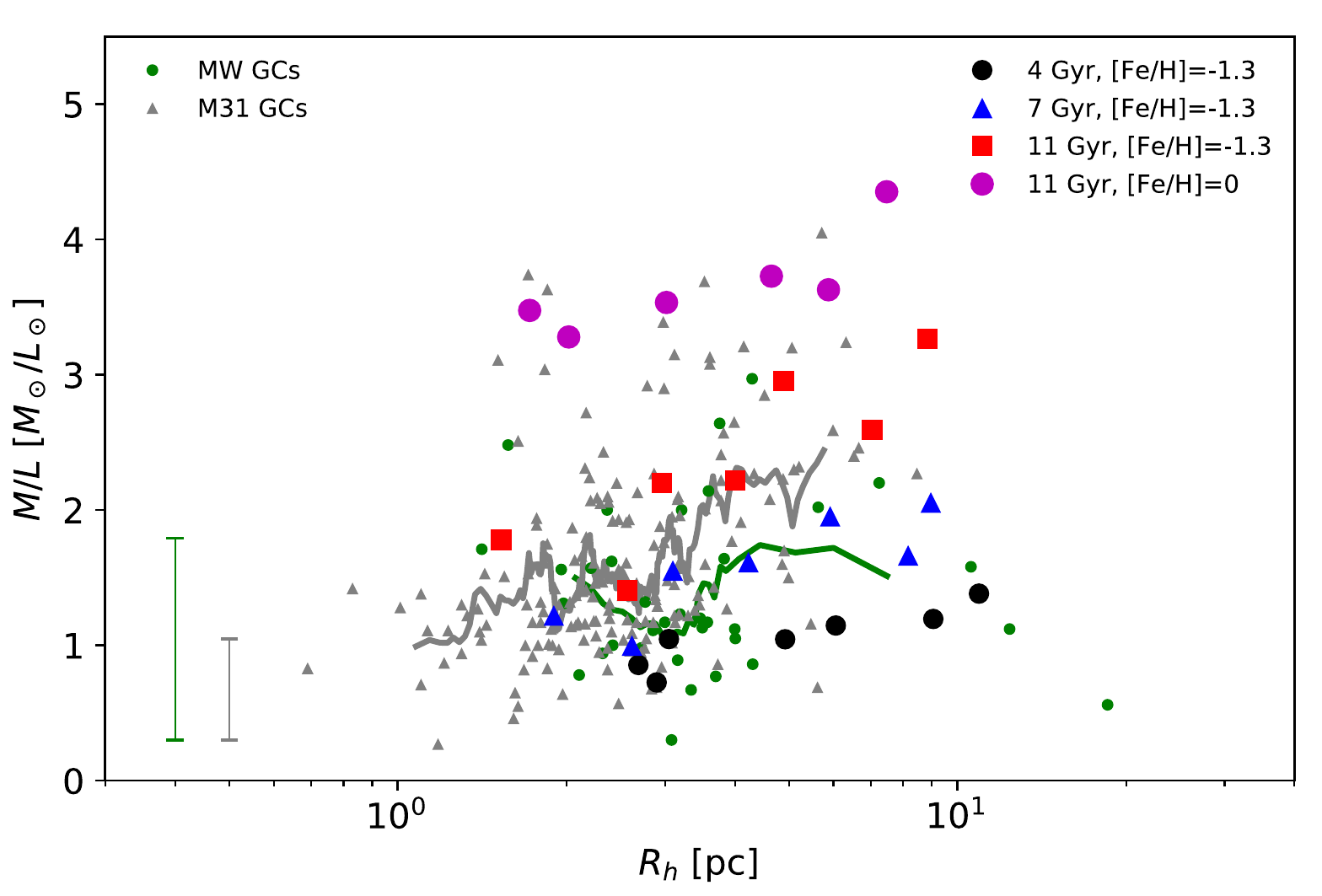}
\caption{Dynamical mass-to-light ratio ($V$-band) as a function of the half-light radius for GCs in the Milky Way (\citealp{MLvdM2005}, green dots) and in M31 (\citealp{Strader2011}, grey triangles). The green and grey solid lines are the sliding means calculated from the data. The models presented in this paper are superimposed and shown as large black circles, blue triangles, red squares, for the metal poor models at 4, 7 and 11 Gyr, respectively, and as magenta circles for metal rich models at 11 Gyr. The average error bars for the observed M/L are indicated on the bottom-left corner. The combination of the different relaxation states of the clusters and the presence of both metal poor and metal rich systems allows to reproduce the parameter space covered by observations and reproduce the observed trend.}
\label{fig:ML_trelh}
\end{figure*}

\section{Comparison with observations}
\label{sec:comparison}

\begin{figure}
\centering
\includegraphics[width=0.5\textwidth]{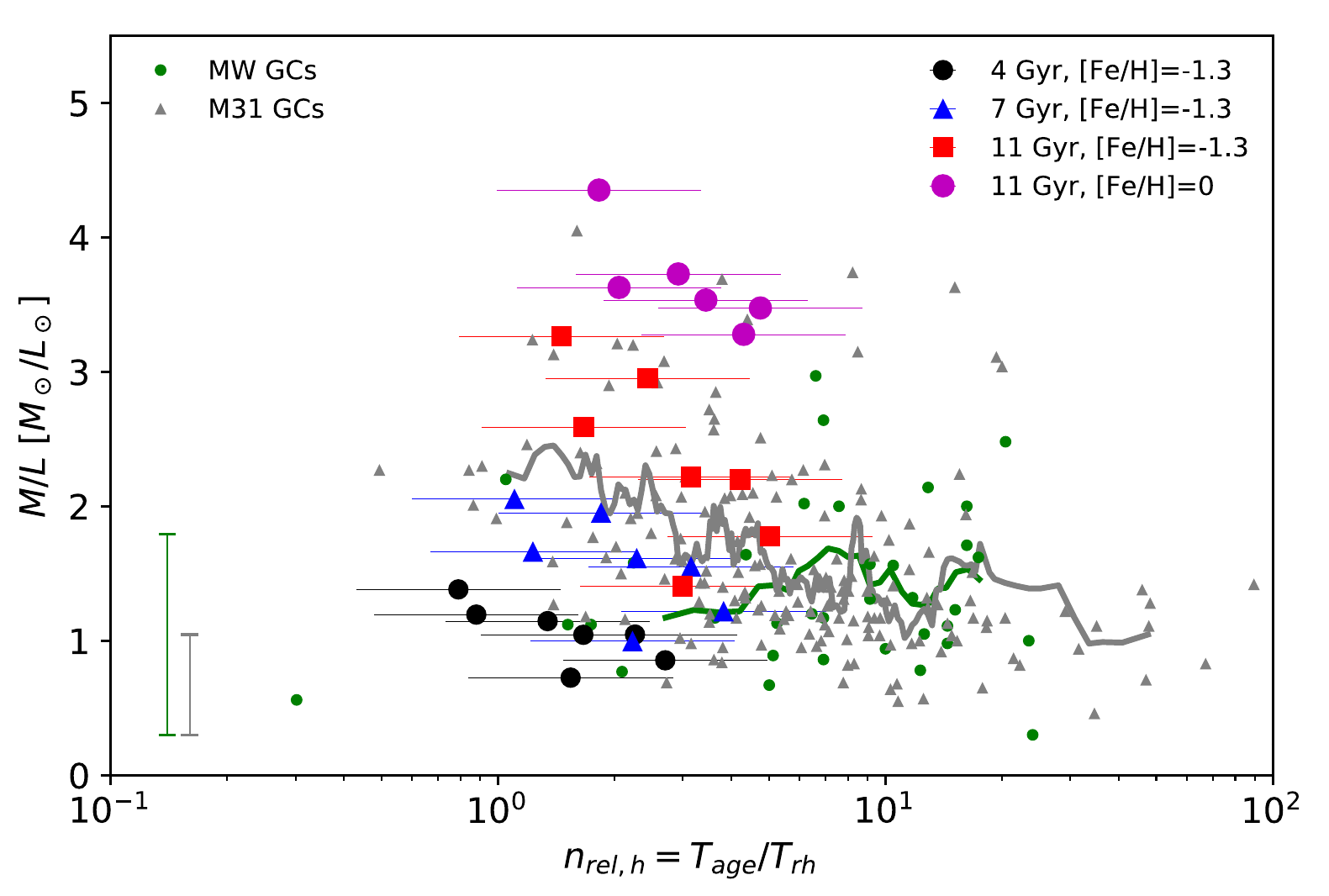}
\caption{Dynamical mass-to-light ratio ($V$-band) as a function of the number of half-mass relaxation times experienced by a cluster, $n_{rel,h}=T_{age}/T_{rh}$, compared to our model prediction. The colour coding is as in Fig.~\ref{fig:ML_trelh}. The green and grey solid lines are the sliding means calculated from the data. The error bars on the relaxation times of the model reflect the uncertainty in the definition of the half-light radius, as discussed in the text. An average age of 8 Gyr is assumed for M31 GCs. Clusters in a more advanced dynamical state (i.e. higher $n_{rel,h}$) display lower mass-to-light ratios fully consistent with our model prediction.}
\label{fig:ML_nrel}
\end{figure}

For a comparison with observation of the relation between relaxation states and mass-to-light ratio from our simulations, we collect a data set comprising dynamical M/L ratios and present day relaxation times, for GCs in the Milky Way and M31.
\begin{itemize}
\item For Milky Way GCs we use the M/L ratios derived from \cite{WIlson1975} dynamical models in the dataset of \citet{MLvdM2005}, and the relaxation times (both the half-mass $T_{rh}$ and the core relaxation times $T_{rc}$) from \citet{Harris1996} (2010 edition)\footnote{We exclude from the sample NGC 6535, due to its high M/L and unusual bottom-light mass function, possibly due to the presence of a massive central black hole or to an extreme dynamical evolution (e.g., \citealp{Zaritsky2014,Askar2017}).}.
\item For M31 we use the dynamical M/L from \citet{Strader2011} and we calculate the half-mass relaxation time $T_{rh}$ with equation 2-63 from \citet{Spitzer1987}, using the total mass estimate and half-light radius in this paper and an estimate of the average stellar mass in the systems of 0.3~$M_\odot$. This is consistent with \citet{Strader2011} and with eq. 2 in \citet{Bianchini2016b} that has been used to calculate the relaxation times for our set of simulations.
\end{itemize}

In Fig. \ref{fig:ML_trelh}, we first compare the observed global M/L to the one predicted by our models as a function of the half-light radius. Our simulations are able to cover the parameter space sampled by the data, explaining variations in the M/L up to a factor of 4, with the presence of clusters with different relaxation states, different metallicities and different ages.

In Fig. \ref{fig:ML_nrel}, we compare the observed relation between the global M/L and the relaxation state of the clusters to the one predicted by our models. Since it is not possible to extract the core relaxation time from the data set of M31, we will use the half-mass relaxation time $T_{rh}$ and define the relaxation state as $n_{rel,h}=T_{age}/T_{rh}$. We wish to point out that the value reported for the relaxation times can be affected by strong uncertainties due to the difficulties in determining a solid estimate of the half-light radius and the dependence $T_{rh}\propto R_h^{3/2}M^{1/2}$. Moreover, the half-light radius is usually measured with different techniques for extragalactic and Galactic GCs, the former relying on unresolved photometry and the latter on (partially-)resolved photometry. Since it is challenging to reliably assess observationally these sources of biases and uncertainties, we decide to take them into account in our models. We calculate an uncertainty in the relaxation time\footnote{The estimate of the relaxation time of our simulations rely on the half-light radius derived from the projected surface brightness profile.} by considering the effect of increasing or decreasing the half-mass radius by a factor of 1.5. Finally, since the ages for M31 GCs are not defined, we assume an average age of 8 Gyr. For the Milky Way GCs we use the ages given by \citet{MLvdM2005}.

The data show a trend of decreasing mass-to-light ratios for dynamically older clusters, in agreement with the finding of Section \ref{sec:globalML} in which more relaxed systems (i.e. systems with lower relaxation times, at a given age) have lost a larger number of massive dark remnants because of dynamical ejections due to the more frequent gravitational encounters (see Fig. \ref{fig:remn_vs_MS}, panel B).

\section{Discussion}
\label{sec:discussion}

In the previous sections we showed that the mass-to-light ratios correlate with the relaxation states of a cluster, with more relaxed clusters showing lower mass-to-light ratios. At variance with previous studies showing that dynamical evolution causes a depletion of low-mass stars and therefore leads to a decrease of the mass-to-light ratio (e.g. \citealp{BaumgardtMakino2003,Kruijssen2008}), we showed that for our simulations the primarily driver of the decrease of the M/L is the ejection of dark remnants. Our finding can be understood in light of the choice of a initial \citet{Kroupa2001} mass function with upper stellar mass of 150 $M_\odot$, rather than 15 $M_\odot$ as in \citealp{BaumgardtMakino2003}, affecting the number of dark remnants created. This indicates that the number of remnants initially retained is a fundamental factor that can affect the values of M/L in GCs, and therefore needs to be explored in greater details.

Observationally, \citet{Strader2011} further showed that the mass-to-light ratios of M31 GCs depends on their metallicity: metal rich clusters have a lower mass-to-light ratio. This trend is in disagreement with simple stellar population models. Several explanations have been proposed, including the need for a different IMF for metal poor and metal rich clusters (\citealp{Strader2009}), a metallicity and density dependent top heavy IMF (\citealp{Zonoozi2016,Haghi2017}) or observational biases due to mass segregation and the use of integrated-light properties (\citealp{ShanahanGieles2015}).

Fig. \ref{fig:ML_nrel} shows that in our simulations, metal rich clusters do not show significant differences in relaxation times from those of metal poor clusters with same initial structural parameters. Therefore we cannot explain the metallicity-M/L trend observed by \citet{Strader2011} for M31 GCs, since metal rich clusters have a systematically higher M/L for every given relaxation state. However, we could speculate that metal rich cluster formed preferentially with shorter relaxation times (e.g. more dense), so that their lower mass-to-light ratios could be explained. Alternatively, since stellar feedbacks are more damaging with increasing metallicity (e.g. \citealp{Dib2011}), the surviving metal rich clusters would preferentially be the densest systems, therefore the one characterized by shorter relaxation times.
In both cases, the onset of energy equipartition and mass segregation would happen on a shorter timescale, making the systems dynamically older than the corresponding metal poor clusters. Thus, in accordance with our findings, dynamical evolution alone could explain their lower M/L.

A signature of a dependence of the relaxation time on the metallicity of the clusters is present in the data of \citet{Strader2011}, where GCs with higher metallicity show a lower relaxation time, on average. However, we should be cautious about this trend due to possible observational effects biasing the measurements of the half-light radii (and therefore of the relaxation times) of metal rich clusters (i.e., the combined effect of stellar evolution and mass segregation can bias the cluster size measurements, giving lower values for the half-light radii for metal rich clusters; see for example  \citealp{Sippel2012,ShanahanGieles2015}).

A hint that relaxation processes are key in explaining the observations, comes from the fact that the discrepancy between the data and simple population models is stronger for clusters with lower mass and with shorter relaxation time \citep{Strader2011}. Indeed, systems with shorter relaxation times would be the one more affected by dynamical evolution effects. Finally, the observed trend of increasing M/L for increasing cluster masses clearly reported for M31 GCs (\citealp{Strader2011}) and NGC 5128 GCs (\citealp{Taylor2015}), and marginally evident for Milky Way GCs (likely because of the lack of very massive clusters, \citealp{Kruijssen2008}), can be taken as an additional indication of the importance of the relaxation states (given the dependence of the relaxation time $T_{rh}\propto R_h^{3/2}M^{1/2}$, more massive clusters have longer relaxation times and therefore higher M/L ratios).

As a final remark, we note that our models are not able to explain the very high M/L values ($M/L>4$ $M_\odot/L_\odot$) observed for some GC in M31 and NGC 5128 (\citealp{Taylor2015}) or for the MW GC NGC 6535 \citep{Zaritsky2014,Askar2017}.

\section{Conclusions}
\label{sec:conclusions}

The internal structure and dynamics of GCs are strongly shaped by two-body interactions that bring the systems toward a state of partial energy equipartition causing mass segregation. In a previous study we related the state of energy equipartition reached by a cluster to its relaxation state (\citealp{Bianchini2016b}). Here, using the same set of dynamical cluster simulations, with a wide range of concentration, binary fraction, initial particle number and metallicity, we explored the relation between the relaxation states and the mass-to-light ratio of a cluster. Since all our simulations experienced similar tidal interactions (comparable to halo clusters at 9-10 kpc from the centre of the Milky Way), they are ideal to study the effects connected to two-body interactions alone and the onset of energy equipartition. We summarize in the following our main results:

\begin{itemize}
\item \textit{Shape of M/L profiles:} A natural consequence of the onset of energy equipartition is the sinking of massive stars into the central regions of a cluster while low-mass stars move outwards. This effect causes a radial variation of the mass-to-light ratio. We find that, apart from an age dependence due to stellar evolution, the precise shape of the M/L profile depends on the relaxation state of the cluster, characterized by the number of relaxation times experienced by the stellar system, $n_{rel}=T_{age}/T_{rc}$. Dynamically younger clusters show a central peak (up to $\simeq25$ $M_\odot/L_\odot$) that progressively vanishes for more dynamically relaxed clusters. This peak is due to the retention of dark remnants that in turn strongly depends on the number of gravitational encounters experienced by these objects. Since dynamically older cluster have ejected their dark remnants, they exhibit a centrally flat M/L profile, with a slight increase in the outer parts due to low mass stars. Interestingly, all the analyzed profiles show a common minimum at about 0.2-0.3 half-light radii.
\item \textit{Global values of M/L:} The global values of M/L (often referred to as dynamical M/L) also show a clear dependence on the dynamical state, with more relaxed clusters showing a lower mass-to-light ratio. In particular, the M/L is found to relate tightly with the number of relaxation times that a cluster has experienced, $M/L\propto n_{rel}^{-1}$, and with the parameter indicating the state of energy equipartition, $M/L\propto m_{eq}$. At given age, the effects of relaxation state alone can explain a variation of the M/L up to a factor of $\simeq3$.
Our simulations show that the dynamical variation of the M/L is mainly driven by the dynamical ejection of dark remnants rather than by the escape of low-mass stars. This offers the possibility of constraining the number of dark remnants present in a clusters based on accurate measurements of the dynamical mass-to-light ratio.
\item \textit{Comparison with observations:} We compared the results of our models to available observations of Milky Way and M31 GCs. Including simulations with different metallicities as well as simulations in different relaxation states, we are able to explain the observed variations of the M/L up to a factor of $\simeq4$. This comparison confirms the importance of the relaxation states in shaping the internal properties of a GC. Note that our models are not able to explain high M/L ($>4$ $M_\odot/L_\odot$) observed for some extragalactic GCs. Additional dynamical ingredients, such as disk shocking or strong tidal interactions, could possibly mitigate these discrepancies, without the need of invoking IMF variations or presence of dark matter.
\item \textit{Implication for dynamical modelling:} In this work we showed that the assumption of constant M/L does not hold in GCs. Especially, even dynamically young cluster can show a strong radial variation of the M/L profile, exhibiting a strong central peak. When carrying out a dynamical model, one could use the relaxation state of a cluster to assign a corresponding physically motivated M/L profile. This can help minimizing the well-known mass-anisotropy degeneracy and, for example, help disentangling the presence/absence of a dark component (e.g, central massive black holes such as intermediate-massive black holes or dark matter halo). In particular, note that the stellar systems where intermediate-mass black holes have been searched for, are the most massive compact stellar systems (most massive GCs, nuclear star clusters, and ultra-compact dwarfs). These systems are the one with longer relaxation times and therefore are dynamically younger. Our work thus suggests they are the one that can be most strongly impacted by a central peak of the M/L profile due to the retention of stellar dark remnants. Such a peak is strongly degenerate with the presence of an intermediate-mass black hole. 
\end{itemize}

\section*{Acknowledgments}
PB acknowledges financial support from a CITA National Fellowship. PB, GvdV and AS acknowledge support by Sonderforschungsbereich SFB 881 "The Milky Way System" (subproject A7 and A8) of the German Research Foundation (DFG). We thank the referee for useful comments and Anna Lisa Varri for interesting discussions.


\bibliographystyle{mnras} 
\bibliography{biblio} 

\appendix
\section{Fits to the M/L profiles}
\label{appA}
We provide fits for the M/L profiles presented in Sect. \ref{sec:darkremn}. The fitting function used is a third-order polynomial function in logarithmic scale
\begin{equation}
\label{fitpoly}
\begin{array}{ll}
M/L=a X^3 + b X^2 + c X + d,\\
X=\log(R/R_h).
\end{array}
\end{equation}
Note that the fitting function is not physically motivated, but is meant to provide a good description of the M/L profiles and capture the main features: central peak for the dynamical young clusters, a slight increase in the outer parts for dynamically old clusters and the minimum in the intermediate region common to all simulations.

\begin{figure*}
\centering
\includegraphics[width=0.9\textwidth]{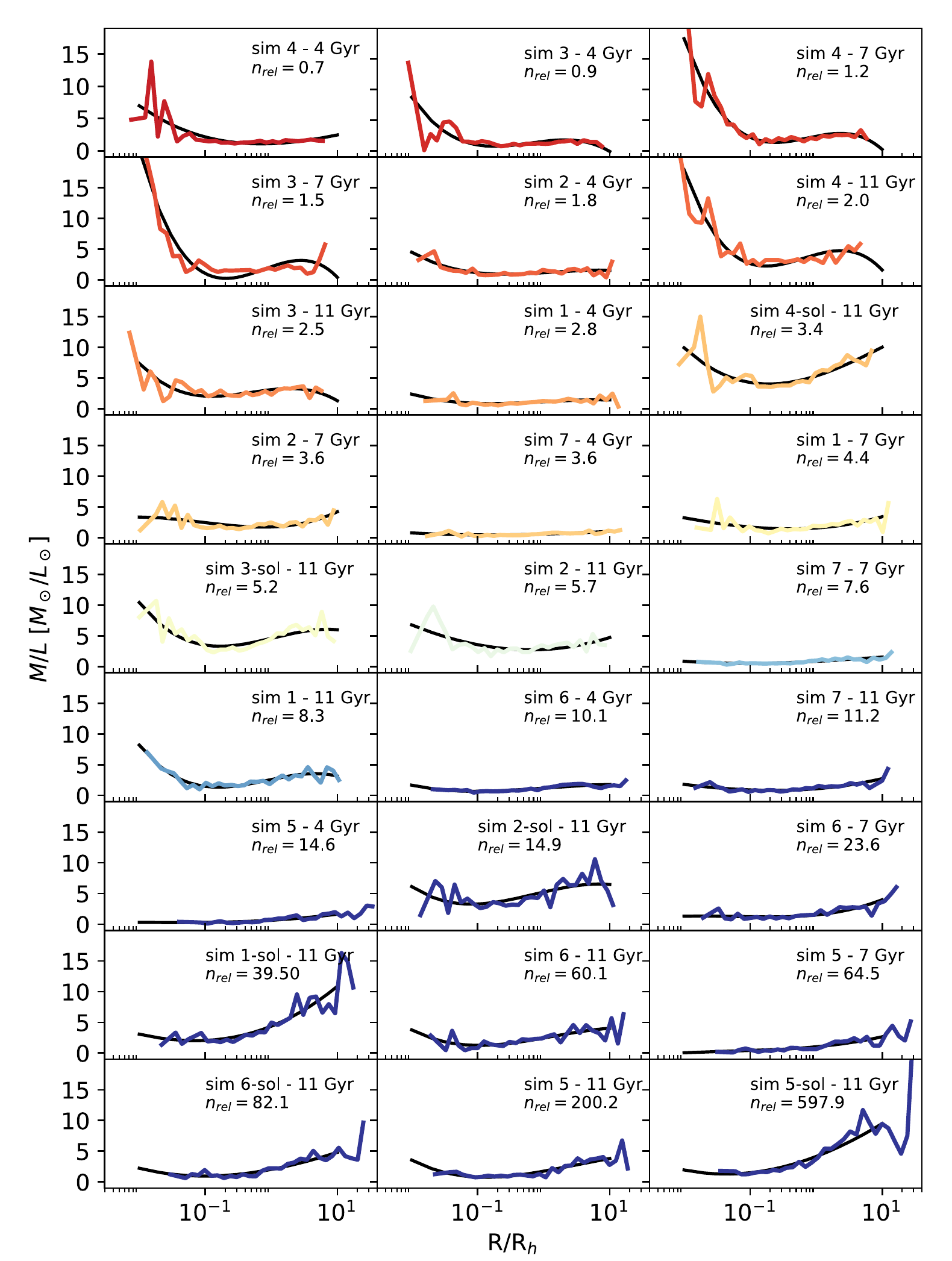}
\caption{M/L profiles for all our simulations with corresponding fit from eq. \ref{fitpoly}. The profiles are colour-coded according to their relaxation state from Table \ref{tab:2} (from top to bottom: redder profiles correspond to dynamically young clusters and bluer profiles to dynamically oder ones).  }
\label{fig:fits}
\end{figure*}

We report in Table \ref{tab:fits} the best fit parameters and the radial position of the minimum for each profile calculated from the fit. We find that the average minimum is $R_{min}=0.22$ $R_h$. Finally, in Fig. \ref{fig:fits} we show all the profiles and corresponding fits, ordered according to their relaxation state $n_{rel}=T_{age}/T_{rc}$.

\begin{table}
\tabcolsep=0.15cm
\begin{center}
\caption{Best fit parameters of the polynomial fitting function given by equation \ref{fitpoly} to the M/L profiles. The last column reports the radial position of the minimum in unit of the half-light radius.}
\begin{tabular}{lccccc}
\hline\hline
 & a& b& c& d& $R_{min}/R_h$\\
\hline
4 Gyr&&&&&\\
\hline
Sim 1&-0.29 & 0.01 & 0.58 & 1.14 & 0.16\\
Sim 2&-0.50 & 0.18 & 0.70 & 1.17 & 0.26\\
Sim 3&-1.78 & -1.14 & 1.24 & 1.59 & 0.18\\
Sim 4&-0.23 & 1.20 & 0.39 & 1.09 & 0.70\\
Sim 5&0.04 & 0.32 & 0.63 & 0.64 & 0.04\\
Sim 6&-0.25 & 0.02 & 0.81 & 1.19 & 0.10\\
Sim 7&-0.02 & 0.16 & 0.31 & 0.57 &0.15\\
\hline
7 Gyr&&&&&\\
\hline
Sim 1&0.11 & 1.00 & 0.73 & 1.56 & 0.40\\
Sim 2&0.50 & 1.60 & 0.39 & 1.73 & 0.74\\
Sim 3&-4.02 & -1.44 & 3.71 & 2.16 & 0.21\\
Sim 4&-3.10 & -1.17 & 2.11 & 2.41 & 0.24\\
Sim 5&0.11 & 0.50 & 1.02 & 1.02 & $-$\\
Sim 6&0.30 & 1.09 & 1.08 & 1.51 & 0.20\\
Sim 7&-0.06 & 0.17 & 0.58 & 0.83 & 0.08\\
\hline
11 Gyr&&&&&\\
\hline
Sim 1&-1.41 & -0.29 & 2.24 & 2.59 & 0.16\\
Sim 2&0.05 & 1.37 & 0.54 & 2.76 & 0.63 \\
Sim 3&-1.54 & -1.44 & 1.16 & 3.11& 0.13\\
Sim 4&-3.46 & -2.00 & 2.94 & 4.12 & 0.17\\
Sim 5&-0.44 & 0.55 & 1.95 & 1.75 & 0.14\\
Sim 6&-0.56 & 0.19 & 1.95 & 2.45 & 0.11\\
Sim 7&0.06 & 0.73 & 0.86 & 1.06 & 0.23\\
Sim 1-sol&0.14 & 2.08 & 4.28 & 4.38 & 0.07\\
Sim 2-sol&-0.92 & -0.37 & 2.50 & 5.24 & 0.08 \\
Sim 3-sol&-1.29 & 0.14 & 2.55 & 4.58 & 0.17\\
Sim 4-sol&-0.51 & 1.82 & 3.38 & 5.33 & 0.19\\
Sim 5-sol&-0.07 & 1.36 & 4.07 & 4.14 & 0.04\\
Sim 6-sol&-0.11 & 0.85 & 2.04 & 5.24 & 0.10\\

\hline
\end{tabular}
\label{tab:fits}
\end{center}
\end{table}

\end{document}